\documentclass[final,5p]{elsarticle}
\pdfoutput=1

\usepackage{amsmath,amssymb,amsthm}
\usepackage{amsfonts,mathrsfs}
\usepackage{lineno,hyperref}
\usepackage{bbm,tikz,tikzpagenodes}

\usepackage[mathcal]{eucal}

\usepackage{microtype}

\usepackage{tikz}

\DeclareMathAlphabet\Euler{U}{eur}{m}{n}

\journal{arXiv}

\newcommand{\nc}{\newcommand}

\nc{\hot}{\,\hat\otimes\,}
\nc{\ot}{\otimes}
\nc{\op}{\oplus}
\nc{\ol}{\overline}
\nc{\un}{\underline}

\nc{\mc}{\mathcal}
\nc{\ms}{\mathsf}
\nc{\mf}{\mathfrak}
\nc{\mb}{\mathbf}
\nc{\bb}{\mathbb}
\nc{\mr}{\mathrm}

\nc{\al}{\alpha}
\nc{\bet}{\beta}
\nc{\eps}{\epsilon}
\nc{\veps}{\varepsilon}
\nc{\del}{\delta}
\nc{\Del}{\Delta}
\nc{\ga}{\gamma}
\nc{\Ga}{\Gamma}
\nc{\ka}{\kappa}
\nc{\la}{\lambda}
\nc{\om}{\omega}
\nc{\si}{\sigma}
\nc{\Si}{\Sigma}
\nc{\Ups}{\upsilon}
\nc{\vphi}{\varphi}
\nc{\pa}{\partial}

\nc{\ii}{{\boldsymbol \imath}}
\nc{\ee}{{\mb e}}

\nc{\id}{\mathrm{id}}
\nc{\mfgr}{\mathrm{gr}}

\nc{\Ug}{U\mathfrak{g}}
\nc{\Ub}{U\mathfrak{b}}

\nc{\ud}{\underline}
\nc{\tl}{\tilde}
\nc{\wt}{\widetilde}
\nc{\wh}{\widehat}

\nc{\End}{\mathrm{End}}
\nc{\Ext}{\mathrm{Ext}}
\nc{\Hom}{\mathrm{Hom}}
\nc{\Ima}{\mathrm{Image}}
\nc{\Ind}{\mathrm{Ind}}
\nc{\Ker}{\mathrm{Ker}}
\nc{\betm}{\mathrm{RHom}}
\nc{\Sym}{\mathrm{Sym}}
\nc{\Str}{\mathrm{Str}}

\nc{\C}{\mathbb{C}}
\nc{\N}{\mathbb{N}}
\nc{\Z}{\mathbb{Z}}

\nc{\lan}{\langle}
\nc{\ran}{\rangle}

\nc{\mfg}{\mf{g}}
\nc{\mfa}{\mf{a}}

\nc{\ua}{{\uparrow}}
\nc{\da}{{\downarrow}}
\nc{\uda}{{\ua\!\da}}

\nc{\cui}[1]{\ms{c}_{#1\ua}}
\nc{\cdi}[1]{\ms{c}_{#1\da}}
\nc{\cdui}[1]{\ms{c}^\dag_{#1\ua}}
\nc{\cddi}[1]{\ms{c}^\dag_{#1\da}}
\nc{\nmui}[1]{\ms{n}_{#1\ua}}
\nc{\nmdi}[1]{\ms{n}_{#1\da}}

\nc{\cu}{\ms{c}_{\ua}}
\nc{\cd}{\ms{c}_{\da}}
\nc{\cdu}{\ms{c}^\dag_{\ua}}
\nc{\cdd}{\ms{c}^\dag_{\da}}
\nc{\nmu}{\ms{n}_{\ua}}
\nc{\nmd}{\ms{n}_{\da}}

\newcommand{\minus}{\scalebox{0.75}[1.0]{$-$}}

\nc{\eqa}[1]{\begin{align}#1\end{align}}
\nc{\eqn}[1]{\begin{align*}#1\end{align*}}
\nc{\eq}[1]{\begin{equation}#1\end{equation}}
\nc{\spl}[1]{\begin{equation}\begin{aligned}#1\end{aligned}\end{equation}}
\nc\el{\nonumber\\}
\nc\nn{\nonumber}

\nc{\qu}{\quad}
\nc{\qq}{\qquad}

\nc{\ket}[1]{|#1\rangle}
\nc{\vac}{|0\rangle}

\nc{\ident}[1]{\vspace{0.1cm}\noindent{\bf #1}}

\nc{\Osc}{\Euler{Osc}}

\numberwithin{equation}{section}

\nc{\red}{\color{red}}
\nc{\blu}{\color{blue}}
\nc{\gre}{\color{green!50!black}}

\nc{\sfrac}[2]{{\textstyle\frac{#1}{#2}}}
\nc{\half}{\sfrac{1}{2}}
\nc{\ihalf}{\sfrac{\ii}{2}}
\nc{\quarter}{\sfrac{1}{4}}

\nc{\alg}[1]{\mathfrak{#1}}
\nc{\rep}{\rho}

\begin{document}

\begin{frontmatter}

\title{An algebraic approach to the Hubbard model}
  
\author{Marius de Leeuw}
\address{The Niels Bohr Institute, University of Copenhagen \\ Blegdamsvej 17, DK-2100 Copenhagen \O , Denmark}
\ead{deleeuwm@nbi.ku.dk} 
 
\author{Vidas Regelskis}
\address{Department of Mathematics, University of Surrey \\ Guildford, Surrey, GU2 7XH, United Kingdom}
\ead{v.regelskis@surrey.ac.uk} 
   
\begin{abstract}
We study the algebraic structure of an integrable Hubbard--Shastry type lattice model associated with the centrally extended $\alg{su}(2|2)$ superalgebra. This superalgebra underlies Beisert's AdS/CFT worldsheet R-matrix and Shastry's R-matrix. The considered model specializes to the one--dimensional Hubbard model in a certain limit. We demonstrate that Yangian symmetries of the R-matrix specialize to the Yangian symmetry of the Hubbard model found by Korepin and Uglov. Moreover, we show that the Hubbard model Hamiltonian has an algebraic interpretation as the so-called secret symmetry. We also discuss Yangian symmetries of the A  and B models introduced by Frolov and Quinn.
\end{abstract}
 
   
\end{frontmatter}


\section{Introduction}

\begin{tikzpicture}[remember picture,overlay]
\node at ([xshift=-2.7cm,yshift=-1.5cm]current page.north east) {DMUS-MP-15/13};
\end{tikzpicture} \vspace{-1.25em}

Exactly solvable models of strongly correlated electrons are of great importance in theoretical condensed matter physics. For instance, they play a prominent role in understanding high-$T_c$ superconductivity. The key example of such a model is the one--dimensional Hubbard model introduced in \cite{Hubbard}. It describes the dynamics of interacting electrons in a one--dimensional lattice that models the conduction band of a solid. Each site in the lattice can have four different states. It can be vacant, occupied by a spin up or down electron or occupied by an electron pair. 

The Hubbard Hamiltonian $\mathcal{H} = \ii \sum_{i} \mc{K}_{ii+1} + \hbar \sum_i \mc{V}_{i}$ for the infinite lattice is written in terms of the usual creation and annihilation operators $\ms{c}^{\dag}_{i\al}$, $\ms{c}_{i\al}$, with $\al=\ua,\da$, of electrons with spin up and down. Here 
\eqn{
\mc{K}_{ii\hspace{-.2mm}+\hspace{-.2mm}1} \!:= \!\!\sum_{\al = \ua,\da} \!(\ms{c}_{i\al}^{\dag}
\ms{c}_{i\hspace{-.2mm}+\hspace{-.2mm}1,\al} \!+\! \ms{c}^\dag_{i\hspace{-.2mm}+\hspace{-.2mm}1,\al} \ms{c}_{i\al} ),\; 
\mc{V}_i \!:= (\ms{n}_{i\ua}\!-\!\half) (\ms{n}_{i\da}\!-\!\half), \\[-2em]
}
where $\ms{n}_{i\al}:=\ms{c}^\dag_{i\al}\ms{c}_{i\al}$ is the number operator, $i\in\N$ enumerates lattice sites, $\ii := \sqrt{-1}$ and $\hbar$ is the coupling constant.

The integrability of the Hubbard model is known since the works of Lieb and Wu \cite{LW}. However, the R-matrix was only found much later by Shastry \cite{Shastry}. 
Given an integrable model, the R-matrix can usually be found from the underlying symmetries of the model. Such symmetries of the Hubbard model were unknown until very recently. 

The Hubbard model was long known to have an exact $\mf{so}(4)=(\mf{su}(2)\op\mf{su}(2))/\Z_2$ symmetry \cite{HL,EKS2} that can be extended to a certain Yangian \cite{UK}. Nevertheless, this was not sufficient to determine Shastry's R-matrix and it was suspected that there are more symmetries underlying the Hubbard model. 

The answer to this question of the full symmetry algebra of the Hubbard model came from an unexpected area -- superstring theory. The R-matrix of the Hubbard model emerges in the prime example of the AdS/CFT correspondence as the worldsheet R-matrix \cite{Be1,MM,Gomez}. It is the intertwining matrix for fundamental representations of the centrally extended $\alg{su}(2|2)$ superalgebra. We will denote this superalgebra by $\alg{g}$. Interestingly, this superalgebra has a non-standard Yangian extension, which is also a symmetry of the R-matrix \cite{Be2}. An exceptional feature of this Yangian is that it has an additional generator, which has no counterpart in $\mf{g}$, called the secret symmetry \cite{MMT}. The complete Yangian symmetry is an infinite--dimensional superalgebra of a novel type \cite{BMdL}.

In this letter we consider an integrable one--dimensional lattice model by identifying each site of the lattice with a fundamental module of $\mf{g}$. This gives an integrable Hubbard--Shastry type model which specializes to the Hubbard model in a certain limit. Such generalized models were first considered in \cite{SW95} and \cite{USW98}. We call this model the general model. Likewise, there are three other parity invariant lattice models that follow from the general model, -- the A and B models of Frolov--Quinn \cite{FQ} and the Essler--Korepin--Shoutens (EKS) model \cite{EKS1}. 

We construct an oscillator representation of $\mfg$, its Yangian extension and the R-matrix, and discuss the integrable structure of the general model and its specializations. In particular we shed more light on the symmetries of the Hubbard model and obtain the spin Yangian symmetry of the Hubbard model \cite{UK} as a certain specialization of the Yangian of $\alg{g}$. Moreover, we show that the Hubbard Hamiltonian is a specialization of the secret symmetry. We also obtain Yangian symmetries of the A and B models.
 
This letter is organized as follows. We first introduce the necessary notation, the setup of the lattice, and the algebra $\alg{g}$. We then construct an oscillator representation of $\alg{g}$ and its Yangian extension, which allows us to compute the R-, Lax- and transfer matrices of the general model. In the remaining sections we discuss links with the A, B and EKS models, the Hubbard model limit, and prove the aforementioned results.


\section{Oscillators and vector spaces}


\paragraph{Oscillator algebra}

We consider a one-dimensional lattice with $L\in\N$ sites that can be occupied by spin-$\half$ particles. Each lattice site is identified with a four-dimensional $\Z_2$-graded vector space $V_i\cong \C^{2|2}$ spanned by vectors
$$
V_i = {\rm span}_\C \{\, \ket{0}_i , \; \ket{\da}_i ,\; \ket{\ua}_i, \; \ket{\uda}_i\,\}
,$$
where $1\le i\le L$ is the index of the site in the lattice. The entire lattice is the $L$-fold tensor product $V := \bigotimes^L_{i=1} V_i$. The vector $\ket{0}_i$ is the vacuum (i.e.\@ a hole at the $i^{th}$ site), $\ket{\ua}_i$ and $\ket{\da}_i$ are spin up and spin down particles, and $\ket{\uda}_i$ is a multi-particle state occupied by a pair of spin up and spin down particles. The grading is 1 for vectors $\ket{\ua}_i$ and $\ket{\da}_i$, and 0 otherwise. 
To describe such a lattice, we introduce the usual free-fermion oscillator algebra generated by $\ms{c}^\dag_{i\al}$ and $\ms{c}_{i\al}$ with $1\le i,j\le L$ and $\alpha,\beta = \ua,\da$, that satisfy the standard (anti)commutation relations
$$
 \{\ms{c}_{i\al} ,\ms{c}_{j\bet}  \} = 0, \qu
\{\ms{c}^\dag_{i\al} ,\ms{c}^\dag_{j\bet}  \} = 0, \qu
\{\ms{c}^\dag_{i\al} ,\ms{c}_{j\bet}  \} = \ms{1}\, \delta_{ij}\delta_{\al\bet}. 
$$
We will denote the universal enveloping algebra of a single copy the oscillator algebra by $\Osc_i$. Likewise, $\Osc_{ij(k...)}$ will denote the universal enveloping algebra of $ij(k...)$ copies of the oscillator algebra. For ease of notation, we identify the unit $\ms{1}$ of the oscillator algebra with the unit $1$ of the ground field $\C$. Upon identification
\eq{
\ket{\al}_i := \ee^{\ii \phi_\al i} \ms{c}^\dag_{i\al}\vac_i, \qu
\ket{\uda}_i := \ee^{\ii (\phi_\ua + \phi_\da) i} \ms{c}^\dag_{i\ua}\ms{c}^\dag_{i\da}\vac_i \label{eq:vectors}
}
and requiring $\ms{c}_{i\al} \vac_i = 0$, the space $V_i$ becomes a left $\Osc_i$- module; here $\ii :=\sqrt{-1}$ and $\phi_\al\in\C$ are arbitrary phases. 


\paragraph{Matrix representation}

Let $E_{ab}\in\End (V_i)$, with $1\le a,b\le 4$, be the usual graded matrix units satisfying
\eqn{
[E_{ab}, E_{cd}] = \del_{bc} E_{ad} - (-1)^{(\bar a+ \bar b)(\bar c +\bar d)} \del_{ad} E_{cb} , \qq \\
(E_{ab} \ot E_{cd})(E_{ij} \ot E_{kl}) = (-1)^{(\bar c +\bar d)(\bar i + \bar j)} E_{ab}E_{ij} \ot E_{cd}E_{kl} ,
}
where $\bar 1=\bar 4=0$ and $\bar 2=\bar 3=1$ is the grading. A matrix representation of the oscillator algebra is given by the map $\pi_i : \Osc_i \to \End (V_i)$ so that 
\eqa{
&\cdi{i} \mapsto \ee^{\ii \phi_\da i} (E_{12} - E_{34}) , \;&& \cui{i} \mapsto \ee^{\ii \phi_\ua i} (E_{13} + E_{24}) , \el
&\cddi{i} \mapsto \ee^{-\ii \phi_\da i} (E_{21} - E_{43}) , && \cdui{i} \mapsto \ee^{-\ii \phi_\ua i} (E_{31} + E_{42}) , \el
& \nmdi{i} \mapsto E_{22} + E_{44},  && \nmui{i} \mapsto E_{33} + E_{44} . 
}
The additional phases $\phi_\al$ in the expressions above are required to make the Hamiltonian and the R-matrix of the model manifestly invariant (i.e.\ without additional twists or similarity transformations) under the standard spin and charge symmetries. This will be explained in detail in Section 4. For subsequent reference we also set $(\minus)^\da := 1$ and $(\minus)^\ua := -1$, $(\minus)^\uda := -1$.


\section{Superalgebra}


\paragraph{Superalgebra} 

We recall the definition of $\alg{g}$ due to \cite{Be1}. By $[\cdot,\cdot]$ we will denote the graded commutator $[a,b]=ab-(-1)^{\bar a \bar b}ba$ for any elements $a$ and $b$ in the superalgebra. The bar $^{-} : \mfg \to \Z_2$ denotes the degree of the element under the $\Z_2$-grading. We will call grade 0 elements bosonic. Likewise, grade 1 elements will be called fermionic.

The centrally extended superalgebra $\alg{su}(2|2)$ is generated by elements $E_{a}$, $F_{a}$, $H_{a}$, with $a=1,2,3$, and central elements $P$ and $K$ subject to the following defining relations:
\spl{
& [H_a,E_b] = A_{ab} E_b , \qu [H_i,F_j] = -A_{ab} F_b , \\
& [E_a,F_b]=\del_{ab} D_a H_a , \qu [E_1,E_3] = 0 , \qu [F_1,F_3] = 0 , \\
& [E_{c},[E_c,E_2]] = 0 , \qu [F_c,[F_{c},F_2]] = 0 , \\
& [[E_1,E_2],[E_3,E_2]]=P , \qu [[F_1,F_2],[F_3,F_2]]=K 
}
for $a,b=1,2,3$ and $c=1,3$. Here $D={\rm diag}(1,-1,-1)$ is the normalization matrix and $A$ is the Cartan matrix given in the appendix. Dynkin nodes 1 and 3 are bosonic; Dynkin node 2 is fermionic. Consequently, the $\Z_2$-grading is $1$ for $E_2$, $F_2$ and is $0$ otherwise.


\paragraph{Local oscillator representation}

Fix a complex number $\hbar$ (this will be the coupling constant of the model). To each site we associate a spectral parameter $u_i$, so that the lattice is parametrized by $\vec u = (u_1,u_2,...,u_L)$ and $\hbar$. 

To obtain a representation of $\mf{g}$ we introduce a function $x(u)$ and parameters $x^\pm_i$ defined by \cite{BeSmat}
\eqa{\label{x(u)}
\! x(u_i) := \frac{1}{2}\left( u_i + \sqrt{u^2_i-4}\right)\!, \;\;
x^\pm_i := x(u_i\pm\half \hbar)^{\pm1} .
}
The parameters $x^\pm_i$ are Zhukovsky variables satisfying
$$
x^+_i + \frac{1}{x^+_i} - x^-_i - \frac{1}{x^-_i} = \hbar, \qu x^+_i + \frac{1}{x^+_i} + x^-_i + \frac{1}{x^-_i}  = 2 u_i.
$$
Then, for each site, we introduce the local weights
$$
a_i = \frac{\ga_i}{\sqrt{\hbar}} , \;\;
b_i = \frac{\nu_i^2-1}{\sqrt{\hbar}\,\ga_i} , \;\; 
c_i = \frac{\ga_i}{\sqrt{\hbar}\,x^+_i} , \;\;
d_i = \frac{x^+_i(\nu^{2}_i-1)}{\sqrt{\hbar}\,\nu^{2}_i\ga_i} , 	
$$
where $\nu^2_i = {x^+_i}/{x^-_i}$ and $\ga^2_i = \nu_i(x^+_i \!- x^-_i)$.
Local weights describe the fundamental representation of $\mfg$ at each site.\footnote{The fundamental representation of $\mfg$ is a four-dimensional atypical cyclic (non-highest weight) representation. See sections 2.3 and 2.4 in \cite{Be1} for more details.} 

A local oscillator representation of $\mfg$ is given by the map $\eta_i : \mfg \to \Osc_i$ such that (for $a=1,2,3$)
\spl{ \label{eq:pi_i}
& E_a \mapsto \mc{E}_{i,a} , && F_a \mapsto \mc{F}_{i,a} , && H_a \mapsto \mc{H}_{i,a}, \\
& P \mapsto \mc{P}_i, && K \mapsto \mc{K}_i, && C \mapsto \mc{C}_i,
}
where 
\eqa{
& \mc{E}_{i,1} =\ee^{\ii(\phi_\ua+\phi_\da)i} \cdui{i}\cddi{i} , \;\; \mc{F}_{i,1} =\ee^{-\ii(\phi_\ua+\phi_\da)i} \cdi{i}\cui{i}, \el
& \mc{E}_{i,3} = \ee^{\ii(\phi_\ua-\phi_\da)i}\cdui{i}\cdi{i}, \qu \mc{F}_{i,3} = \ee^{\ii(\phi_\da-\phi_\ua)i}\cddi{i}\cui{i}, \el
& \mc{E}_{i,2} = \ee^{-\ii \phi_\ua i}((a_i-b_i) \, \cui{i}\nmdi{i} + b_i\, \cui{i} ) , \qu \mc{P}_{i} = a_i b_i ,  \label{gens-osc} \\ 
& \mc{F}_{i,2} = \ee^{\ii \phi_\ua i} ((d_i-c_i) \, \cdui{i}\nmdi{i} + c_i\, \cdui{i}), \qu \mc{K}_{i}= c_id_i, \el
&  \mc{H}_{i,1} = \nmui{i}\!+\nmdi{i} \!-\! 1, \qu \mc{H}_{i,3} = \nmdi{i}-\nmui{i}, \el
& \mc{H}_{i,2} = {-}\mc{C}_i - \half \mc{H}_{i,1} -\half \mc{H}_{i,3}, \qu  \mc{C}_i = \half(a_id_i+b_i c_i) \nn
}
are operators in $\Osc_i\subset\Osc_{i...L}$.
The operators with labels $1$ and $3$ form two copies of the $\alg{su}(2)$ algebra. At the $i^{th}$ site, for $\al=\ua,\da$, we have 
$$
\mc{E}_{i,3} \ket{\da}_i = \ket{\ua}_i , \qu \mc{F}_{i,3} \ket{\ua}_i = \ket{\da}_i , \qu \mc{H}_{i,3} \ket{\al}_i = (\minus)^{\al} \ket{\al}_i .
$$
Thus operators with label $1$ realize the spin symmetry of the spin-$\half$ particles. Similarly, operators with label $3$ satisfy 
$$
\mc{E}_{i,1} \vac_i = \ket{\uda}_i , \qu \mc{F}_{i,1} \ket{\uda}_i = \vac_i , \qu \mc{H}_{i,1} \ket{\al}_i = - (\minus)^\al \ket{\al}_i 
$$
for $\al=0,\uda$. This is the charge symmetry of the spin-$\half$ particles. From now on we will refer to these operators using this terminology. The action of fermionic operators (for $\al=\da,\uda$ and $\bet=0,\ua$) is
\eqn{
& \mc{E}_{i,2} \ket{\da}_i = b_i \vac_i , \;\; \mc{E}_{i,2} \ket{\uda}_i = a_i \ket{\da}_i , \;\; \mc{H}_{i,2} \ket{\al}_i = -\half a_i d_i \ket{\al}_i ,\\
& \mc{F}_{i,2} \vac_i = c_i \ket{\ua}_i , \;\; \mc{F}_{i,2} \ket{\da}_i = d_i \ket{\uda}_i , \;\; \mc{H}_{i,2} \ket{\bet}_i = -\half b_i c_i \ket{\bet}_i .
}
These operators are supersymmetries of the model.
 

\paragraph{Global oscillator representation}

Let us extend the representation \eqref{gens-osc} to the entire lattice. To achieve this, we need the tensor product describing the lattice $V= \bigotimes^L_{i=1} V_i$ to be defined over the oscillator algebra $\Osc_{1...L}$. We require $\mc{J}_{i,a}\,\ket{\al}_j = \pm\, \ket{\al}_j\,\mc{J}_{i,a}$ if $i\ne j$ for all $\mc{J}_{i,a}$ in \eqref{gens-osc} and $\al \in \{ 0, \ua, \da, \uda \}$. Here the sign is `$-$' if both $\mc{J}_{i,a}$ and $\ket{\al}_j$ are fermionic, and `$+$' otherwise. 
Following the braided Hopf algebra structure of $\mfg$ introduced in \cite{PST}, we define the so-called braiding factors $\nu^{(i)} = \prod_{1\le j \le i} \nu_j$ (we set $\nu^{(0)}  :=  1$), and (for $1\le a \le 3$)
\begin{align} \label{psu22:osc}
& \mc{E}_{a} = \sum_{1\le i\le L} \!(\nu^{(i-1)})^{\delta_{a2}}\, \mc{E}_{i,a}  ,  &&\mc{P} = \sum_{1\le i\le L} \!(\nu^{(i-1)})^{2}\, \mc{P}_{i} , \el
& \mc{F}_{a} = \sum_{1\le i\le L}  \!(\nu^{(i-1)})^{-\delta_{a2}}\, \mc{F}_{i,a} , &&\mc{K} = \sum_{1\le i\le L} \!(\nu^{(i-1)})^{-2}\, \mc{K}_{i} , \el
& \mc{H}_{a} = \sum_{1\le i\le L} \mc{H}_{i,a} , &&\mc{C} = \sum_{1\le i\le L} \mc{C}_{i} .
\end{align}
Notice that the braiding factors $\nu^{(i-1)}$ make the action of the fermionic operators $\mathcal{E}_2,\mathcal{F}_2$ and the central elements $\mathcal{P},\mathcal{K}$ non-local. We will call operators in \eqref{psu22:osc} the global operators. 

The map $\rho : \mfg \to \Osc_{1...L}$ obtained from the one in \eqref{eq:pi_i} by replacing the local operators in \eqref{gens-osc} with their global couterparts in \eqref{psu22:osc} is a global representation of $\mfg$. It is an oscillator algebra realization of the braided coproduct introduced in \cite{PST}.

For any $\ms{c}_{i\al},\ms{c}^\dag_{i\al}\in \Osc_{1...L}$ let
\eq{ 
\sigma: \begin{cases} \ms{c}_{i\al}  \mapsto \ee^{-\ii \phi_\al(L-2i+1)}\ms{c}_{L-i+1,\al} , \\
\ms{c}^\dag_{i\al} \mapsto \ee^{\ii \phi_\al(L-2i+1)}\ms{c}^\dag_{L-i+1,\al} 
\end{cases}
}
be the parity reversal operator. Moreover, require $\si(\Euler{a}_i)=\Euler{a}_{L-i+1}$ for all local parameters (i.e.\@ local weights, Zhukovsky variables, etc.). Then the `opposite' map $\rho' : = \si\circ\rho$ is a realization of the opposite braided coproduct.


\paragraph{Yangian} 

The superalgebra $\alg{g}$ admits a non-conventional Yangian extension $Y(\alg{g})$ \cite{Be2}. To describe its structure we need to introduce local operators $\mc{Q}_{i\al\bet}$ and $\mc{Q}^\dag_{i\al\bet}$ corresponding to all fermionic root vectors of $\mfg$:
\spl{ \label{eq:QQd}
\mc{Q}_{i\ua\ua} &= \mc{E}_{i,2}, && \mc{Q}^\dag_{i\ua\ua} = \mc{F}_{i,2}, \\
\mc{Q}_{i\da\ua} &= [\mc{E}_{i,2},\mc{E}_{i,1}], \qu && \mc{Q}^\dag_{i\da\ua} = [\mc{F}_{i,1},\mc{F}_{i,2}], \\
\mc{Q}_{i\ua\da} &= [\mc{E}_{i,3},\mc{E}_{i,2}], && \mc{Q}^\dag_{i\ua\da} = [\mc{F}_{i,2},\mc{F}_{i,3}], \\
\mc{Q}_{i\da\da} &= [\mc{E}_{i,1},\mc{Q}_{i,2}], && \mc{Q}^\dag_{i\da\da} = [\mc{Q}^\dag_{i,2},\mc{F}_{i,1}].
}
Their global counterparts, $\mc{Q}_{\al\bet}$ and $\mc{Q}^\dag_{\al\bet}$, are defined equivalently. In addition, we define the following non-local quadratic operators
\eqn{
\mc{Z}_{ij} &= \sum_{a=1,3} D_a ( \mc{E}_{i,a} \mc{F}_{j,a} + \mc{F}_{i,a} \mc{E}_{j,a} + \half \mc{H}_{i,a} \mc{H}_{j,a}), \\
\mc{W}^\pm_{ij} &= \sum_{\al,\bet=\ua\!\da} \left( \frac{\nu^{(j-1)}}{\nu^{(i-1)}}  \, \mc{Q}^\dag_{i\al\bet} \mc{Q}_{j\al\bet} \pm \frac{\nu^{(i-1)}}{\nu^{(j-1)}}  \, \mc{Q}_{i\al\bet} \mc{Q}^\dag_{j\al\bet} \right) .
}
For $\mc{J} \in \{\mc{E} ,\mc{F} ,\mc{H} \}$ we then introduce spin and charge Yangian operators (for $a=1,3$)
\begin{align}\label{eq:Yang}
\wh{\mc{J}}_a &= \sum_{1\le j\le L} u_j \mc{J}_{j,a} - \frac{\hbar}{2} \sum_{1\le i< j \le L} \big[ \mc{J}_{i,a} , \mc{Z}_{ij} - \mc{W}^-_{ij} \big].
\end{align}
The Yangian operators corresponding to the remaining generators of $\mfg$ can be obtained using Lie algebra relations. Finally, we introduce an additional Yangian operator
\begin{align}
\label{eq:Secret}
\wh{\mc{B}} &= \sum_{1\le j\le L} \frac{u_j}{a_j d_j + b_j c_j}\, (\mc{H}_{j,3})^2 - \frac{\hbar}{2} \sum_{1\le i<j \le L} \mc{W}^+_{ij} .
\end{align}
The operator $\wh{\mc{B}} $ is called the secret symmetry \cite{MMT}. It has no counterpart in the algebra $\alg{g}$ and extends the Yangian of $\alg{g}$ to a deformed Yangian of $\alg{gl}(2|2)$ \cite{BMdL}.


\paragraph{PPHT} We introduce a partial particle--hole transformation $\ms{P}_\da$, which acts as the identity map on the spin up oscillators, but as involution on the spin down oscillators, and inverts the lattice parameters as follows:
\begin{align} 
\ms{P}_\da:\big\{\ms{c}_{i\ua}, \ms{c}_{i\da},x^\pm_i,\hbar,\phi_\da \big\} \mapsto \big\{\ms{c}_{i\ua},\ms{c}^\dag_{i\da},\frac{1}{x^\mp_i} ,\minus\hbar,\minus\phi_\da \big\}.
\end{align}
This transformation interchanges the spin and charge operators, $\mc{E}_{i,1} \leftrightarrow -\mc{E}_{i,3}$, $\mc{F}_{i,1} \leftrightarrow -\mc{F}_{i,3}$, $\mc{H}_{i,1} \leftrightarrow -\mc{H}_{i,3}$; inverts the elements $\mc{P}_i\to -\mc{P}_i$, $\mc{K}_i\to -\mc{K}_i$, $\mc{C}_i\to -\mc{C}_i$; and acts on the fermionic operators as $\mc{E}_{i,2} \mapsto \ii\mc{E}_{i,2}$, $\mc{F}_{i,2} \mapsto \ii\mc{F}_{i,2}$ due to the presence of $\sqrt{\hbar}$ factors in the local weights.


\section{Integrable structure}


\paragraph{R-matrix}

The R-matrix of an integrable system is determined by its symmetry properties. For a subset $S\subset \{1,...,L\}$ introduce a projector $\veps_{S} : \Osc_{1...L} \to \Osc_{S}$ by
$$
\veps_{S}(\ms{c}_{i\al})= \del_{i\in S} \ms{c}_{i\al} , \;\; \veps_S(\ms{c}^\dag_{i\al}) = \del_{i\in S} \ms{c}^\dag_{i\al}, \;\; \veps_S(\nu_i)=(\nu_i)^{\del_{i\in S}},
$$
where $\del_{i\in S}=1$ if $i\in S$, and $\del_{i\in S}=0$ otherwise. Then, for each of the global operators $\mathcal{J}$ defined in \eqref{psu22:osc} and also for the Yangian operators in \eqref{eq:Yang} and \eqref{eq:Secret}, the oscillator R-matrix is the two-site operator $\mc{R}_{ij}\in \Osc_{ij}$ satisfying the intertwining equation 
\begin{align} \label{eq:intw}
\mc{R}_{ij}\,\veps_{ij}(\mathcal{J}) = (\veps_{ij}\circ\sigma)(\mathcal{J})\, \mc{R}_{ij}.
\end{align}
This equation implies that the oscillator R-matrix is a homogeneous element of degree 0 in $\Osc_{ij}$, and determines it uniquely up to an overall scalar factor (which we set to identity for simplicity), 
\begin{align}\label{eq:Sosc}
\mc{R}_{ij} &= 1 - \half(s^{(4)}_{ij}+s^{(5)}_{ij}) \el
& + \big(\ee^{\ii(\phi_\ua+\phi_\da)(i-j)} s^{(2)}_{ij} \cdui{i}\cddi{i}\cui{j}\cdi{j} \el
& \qq -\ee^{\ii(\phi_\ua-\phi_\da)(i-j)}  s^{(3)}_{ij}\! s^{(2)}_{ji}\cdui{i}\cdi{i}\cddi{j}\cui{j}  + h.c. \big)  \el
& + s^{(2)}_{ij}( \ms{n}_{i\ua} \ms{n}_{i\da} +  \ms{n}_{j\ua} \ms{n}_{j\da}) ( \ms{n}_{i\ua} - \ms{n}_{j\ua}) (\ms{n}_{i\da} -\ms{n}_{j\da}) \nn\\
& + s^{(3)}_{ij}s^{(2)}_{ji} ( \ms{n}_{i\ua} \ms{n}_{i\da} +  \ms{n}_{j\ua} \ms{n}_{j\da}-1) ( \ms{n}_{i\ua}-  \ms{n}_{j\ua}) (\ms{n}_{i\da} -\ms{n}_{j\da}) \nn\\
& + 2 s^{(4)}_{ij}(\ms{n}_{i\ua} -\half)(\ms{n}_{i\da} -\half) + 2s^{(5)}_{ij} (\ms{n}_{j\ua} -\half)(\ms{n}_{j\da} -\half) \nn\\
& + (s^{(3)}_{ij}+s^{(4)}_{ij}+s^{(5)}_{ij}-1) (\ms{n}_{i\ua} - \ms{n}_{i\da})^2\, (\ms{n}_{j\ua} - \ms{n}_{j\da})^2\nn\\
& +\sum_{\al\neq\bet} (\ee^{\ii\phi_\al(i-j)}\ms{c}^\dag_{i\ua} \ms{c}_{j\ua} - \ee^{\ii\phi_\al(j-i)} \ms{c}_{i\ua} \ms{c}^\dag_{j\ua}) \el[-1em]
& \hspace{2.2cm} \times (s^+_{ij}(\ms{n}_{j\bet}-\half)-s^-_{ij} (\ms{n}_{i\bet} -\half)) .
\end{align}
The Boltzmann weights are
\eqa{\label{eq:BolzRmat}
s^{(\pm)}_{ij} &= \bigg(\frac{y^{++}_{ij}}{1-x^-_ix^-_j} \pm \frac{\nu_i\nu_j}{x^-_i - x^+_j}\bigg) \frac{\ga_i\ga_j}{\nu_i^2\nu_j} , \el
s^{(2)}_{ij} &= \frac{(x^+_i - x^-_i)(x^+_j - x^-_ix^-_jx^+_i) } {(x^-_i - x^+_j)(x^+_i - x^-_i x^-_j x^+_i) } , \qu 
s^{(3)}_{ij} = \frac{\nu_j}{\nu_i} \,y^{+-}_{ij} , \el
s^{(4)}_{ij} &= 1- \nu_j \, y^{--}_{ij} , \qu
s^{(5)}_{ij} = 1- \frac{1}{\nu_i}\, y^{++}_{ij} ,
}
where $y^{\al\bet}_{ij}= \dfrac{x^\al_i -x^\bet_j}{x^-_i-x^+_j}$. It is an instructive exercise to check this oscillator R-matrix is unitary, $\mc{R}_{ij}\mc{R}_{ji}=1$, and satisfies the Yang-Baxter equation
\eq{
\mc{R}_{ij}\mc{R}_{ik}\mc{R}_{jk} = \mc{R}_{jk}\mc{R}_{ik}\mc{R}_{ij} . \label{YBE:osc}
}
The transformation $P_\da$ is not a manifest symmetry of $\mc{R}_{ij}$.


\paragraph{Matrix representation}

The graded matrix representation $R(u_i,u_j)$ of the oscillator R-matrix is obtained by setting
\eq{
R(u_i,u_j) := (\pi_i \ot \pi_j) (\mc{R}_{ij}) \in \End (V_i \ot V_j) .
}
For reader's convenience, we have written it explicitly in the appendix. Let $P = \sum_{1\le a,b \le 4} (-1)^{\bar b} E_{ab} \ot E_{ba}$ denote the graded permutation operator on the graded vector space. We have that $R(u,u)=P$ and the unitarity property now reads as $R(u_i,u_j) P R(u_j,u_i) P = I$, where $I$ is the identity matrix. Moreover, $R$ satisfies the graded Yang-Baxter equation in $\End(V_i\ot V_j \ot V_k)$
$$
R_{ij}R_{ik}R_{jk} = R_{jk} R_{ik} R_{ij},
$$
where $R_{ij} = R(u_i,u_j) \ot I$, $R_{jk} = I \ot R(u_j,u_k) $ and $R_{ik} = (I \ot P) R(u_i,u_k) (I \ot P)$.
 
The matrix $R(u_i,u_j)$ is the Beisert's AdS/CFT worldsheet R-matrix found in \cite{BeSmat}. It is equivalent to Shastry's R-matrix up to a similarity transformation and the following identification of the parameters \cite{MM}
\eq{
x^+_i = \ii \frac{\rm a(\theta)}{\rm b(\theta)}\,\ee^{2 \rm h(\theta)} , \qu 
x^-_i = -\ii \frac{\rm b(\theta)}{\rm a(\theta)}\,\ee^{2 \rm h(\theta)} ,  \label{eq:hub-par}
}
with $\rm a^2(\theta) + b^2(\theta) = 1$ and  $\hbar\,\rm a(\theta)\,b(\theta) = 2\ii\sinh 2h(\theta)$. The standard choice is $\rm a(\theta)=\cos \theta$ and $\rm b(\theta)=\sin \theta$.


\paragraph{Lax operator}

In this letter we use Latin letters $i,j$ to index lattice sites, which are quantum spaces of our integrable model. We will use Greek letters $\la,\mu$ to index auxiliary spaces associated with the model. In our setup, auxiliary spaces are isomorphic to the quantum spaces. Thus the Lax operator $\mc{L}_{\mu i}$ is obtained by associating one leg of the oscillator R-matrix with the auxiliary space and mapping it to the matrix representation. By this we mean that
\eq{\label{eq:LviaR}
\mc{L}_{\mu i} := \pi_\mu (\mc{R}_{\mu i}) \in \End (V_\mu) \ot \Osc_i .
}
For any matrices $A$ and $B$, introduce a graded tensor product by
$$
(A\hot B)_{ac,bd} = (-1)^{\bar a (\bar c + \bar d)} A_{ab} B_{cd} .
$$
This allows us to write the Lax operator $\mc{L}_{\mu i}$ in terms of the following decomposition ({\it c.f.}\@ \cite{OW})
$$
\mc{L}_{\mu i} = K_{\mu i}(\mc{L}_{\mu i\ua} \hot \mc{L}_{\mu i\da}) K_{\mu i} + P_{\mu i} (\ms{n}_{i\ua}-\half)(\ms{n}_{i\da}-\half) + \tfrac{1}{4}Q_{\mu i} , 
$$
where $\mc{L}_{\mu i \al}$ is a Lax operator for a fermionic lattice model with two-dimensional sites 
\eq{
\mc{L}_{\mu i \al} = \left(\!\! \begin{array}{cc} 
f_{\mu i} (\ms{n}_{i\al}\!-\!\half)\! -\! \half f_{\mu i}^{-1} & \ee^{\ii \phi_\al i}\ms{c}^\dag_{i\al} \!\! \\[0.3em]
-\ee^{-\ii \phi_\al i} \ms{c}_{i\al} & \!\!\! f_{\mu i} (\ms{n}_{i\al}\!-\!\half)\! +\! \half f_{\mu i}^{-1} 
\end{array} \!\!\right) \!. \!\!\!
}
The matrices $K_{\mu i}$, $P_{\mu i}$ and $Q_{\mu i}$ all have the same form,
\eqn{
K_{\mu i} = \half (k^+_{\mu i} + k^-_{\mu i})\, I + \half(k^+_{\mu i} - k^-_{\mu i} )\,  \si^z \ot \si^z ,
}
where $\si^z$ is the third Pauli matrix. For $P_{\mu i}$ and $Q_{\mu i}$ one needs to replace $k^\pm_{\mu i}$ with $p^\pm_{\mu i}$ and $q^\pm_{\mu i}$, respectively. The matrix elements in the expressions above contain 
\eqn{
(k^+_{\mu i})^2 =  - s^{(2)}_{\mu i} , \qu  (k^-_{\mu i})^2 = s^{(2)}_{ia} s^{(3)}_{\mu i} , \qu f_{\mu i} = - \frac{s^+_{\mu i}}{k^+_{\mu i}k^-_{\mu i}} 
}
and
\eqn{
& q^+_{\mu i} + p^+_{\mu i} =4 \nu_\mu\, (1-s^{(5)}_{\mu i})\, (1-z_{\mu i} \nu_i/x^+_i)  , \\
& q^-_{\mu i} - p^-_{\mu i} =4 \nu_i\, (1-s^{(5)}_{\mu i})\, (1-z_{i\mu} \nu_\mu/x^+_\mu) , \\
& q^-_{\mu i} + p^-_{\mu i} = 4((1-s^{(5)}_{\mu i})\, z_{i\mu} + (1-s^{(4)}_{\mu i}) ) , \\
& q^+_{\mu i} - p^+_{\mu i} = 4(1-s^{(5)}_{\mu i} )\, (1-z_{\mu i})  ,
}
where
\eq{
z_{\mu i} = \frac{(s^+_{\mu i}-s^-_{\mu i})(s^+_{i\mu}+s^-_{i\mu})}{4s^{(2)}_{i\mu}(s^{(5)}_{\mu i}-1)} =\frac{x_i^+}{\nu_i} \frac{1-\nu_\mu^2}{x^+_i x^-_i- \nu_\mu^2}.
}
The inverse of the Lax operator is obtained by a certain inversion of the parameters $x^\pm$ of both auxiliary and quantum spaces, namely
\eq{
\mc{L}_{\mu i}^{-1} = \mc{L}_{\mu i} \big|_{x^\pm \to 1/x^{\mp}} .
}

The Lax operator satisfies the fundamental commutation relation in $\End(V_\mu\ot V_\la) \ot \Osc_i$
\eq{\label{eq:RLL}
R_{\mu\la}(\mc{L}_{\mu i}\ot I) ( I\ot \mc{L}_{\la i}) = ( I\ot \mc{L}_{\la i})(\mc{L}_{\mu i}\ot I)R_{\mu\la} ,
}
which follows from \eqref{YBE:osc}. The intertwining equation \eqref{eq:intw} for bosonic operators in \eqref{psu22:osc} now reads
\eq{
[\mc{L}_{\mu i} , \mc{J}_i] = [\pi_\mu(\mc{J}_\mu),\mc{L}_{\mu i}]. \label{eq:[L,J]}
}
Notice that the commutator on the left hand side acts on the oscillators only, while the commutator on the right hand side deals solely with the matrix structure of $\mc{L}_{\mu i}$. 

Let $\mc{Q}_i\equiv\mc{Q}_{i\al\bet}$ and $\mc{Q}^\dag_i\equiv\mc{Q}^\dag_{i\al\bet}$ be a shorthand notation for fermionic operators in \eqref{eq:QQd}. We have the identity
\eq{ \label{eq:[L,Q]}
\nu_\mu \mc{L}_{\mu i} \mc{Q}_{i} = \mc{Q}_{i} \mc{L}_{\mu i} - \mc{L}_{\mu i} \pi_\mu(\mc{Q}_{\mu}) + \pi_\mu(\mc{Q}_{\mu}) \mc{L}_{\mu i} \nu_i , \\
}
which tells us how to move a local fermionic operator through the Lax operator. A similar identity, with $\nu_\mu$ and $\nu_i$ replaced with their inverses, holds for $\mc{Q}^\dag_i$.


\paragraph{Transfer matrix}

We introduce the monodromy matrix in the usual way 
\begin{align}\label{eq:DefMono}
\mc{T}_\mu : = \mathcal{L}_{\mu1} \cdots \mathcal{L}_{\mu L} \in \End V_\mu \ot \Osc_{1...L} . 
\end{align}
It satisfies a commutation relation in $\End(V_\mu\ot V_\la) \ot \Osc_{1...L}$ similar to \eqref{eq:RLL} 
\eq{\label{eq:RTT}
R_{\mu\la}(\mc{T}_{\mu}\ot I) ( I\ot \mc{T}_{\la}) = ( I\ot \mc{T}_{\la})(\mc{T}_{\mu}\ot I)R_{\mu\la}.
}
For any graded matrix $A\in\End V_i$, introduce a supertrace by $\Str A = \sum_{1\le a\le 4} (\minus)^{\bar a}(A)_{aa}$.
The transfer matrix is defined as the supertrace of the monodromy matrix
\eq{\label{eq:defTrans}
\tau_{\mu} : = \Str_\mu \mc{T}_\mu \in \Osc_{1...L}.
}
Relation \eqref{eq:RTT} implies that $[\tau_{\mu},\tau_{\la}] = 0$. Thus $\tau_{\mu}$ generates a tower of mutually commutative operators that characterize this integrable model. Note that $\tau_\mu \equiv \tau_\mu(u_\mu;\vec{u})$ is a function of the spectral parameters of the auxiliary and all the quantum spaces. 

Let us consider a homogeneous lattice. Each quantum space is now described by the same spectral parameter $u\equiv u_i$ for all $1\le i\le L$. By expanding the transfer matrix in the neighbourhood of $u_\mu = u$, we have
\begin{align}\label{eq:ExpandTrans}
\tau_{\mu}(u_\mu;u) = \Big[\prod_{i=1}^{L-1}\mc{P}_{i i+1}\Big]\Big[1 + (u_\mu-u)\mc{H} +\ldots\Big] ,
\end{align}
where $\mc{H}$ is the Hamiltonian of the general model. We will compute the Hamiltonian explicitly in the next subsection. We now want to focus on the symmetry properties of $\mc{T}_\mu$ and $\tau_\mu$.

Both $\mc{T}_\mu$ and $\tau_{\mu}$ have the usual symmetry properties for the bosonic operators in \eqref{psu22:osc}. They are 
\eq{
[\mc{T}_{\mu} , \mc{J}] = [\pi_\mu(\mc{J}_\mu), \mc{T}_{\mu}] , \qq [\mc{J}, \tau_{\mu}] = 0 . \label{eq:[T,J]}
}
However, for the fermionic operators in \eqref{eq:QQd}, the symmetry relations are of non-abelian type that usually appear in models with quantum deformed symmetries. For the monodromy matrix we find
\eq{
\nu_\mu \mc{T}_\mu \si(\mc{Q}) = \si(\mc{Q}) \mc{T}_\mu - \mc{T}_\mu \pi_\mu(\mc{Q}_\mu) + \pi_\mu(\mc{Q}_\mu) \mc{T}_\mu \nu^{(L)} , \label{eq:[T,Q]}
}
which can be seen from \eqref{eq:[L,Q]}. This implies that
\eq{
\nu_\mu \tau_{\mu} \si(\mc{Q}) = \si(\mc{Q}) \tau_{\mu} - (1-\nu^{(L)})\,\Str_\mu \big(\pi_\mu(\mc{Q}_\mu) \mc{T}_\mu\big) . \label{eq:[t,Q]}
}
The presence of the braiding factors makes the fermionic symmetries non manifest. The operator $\Str_\mu \big(\pi_\mu(\mc{Q}_\mu) \mc{T}_\mu\big)$ is a transfer matrix for a model with twisted boundary conditions. In the $\nu^{(L)}=1$ limit this term does not contribute, and the fermionic operators relates eigenstates of the transfer matrix whose eigenvalues differ by a factor of $\nu_\mu$. Otherwise, the fermionic operators relate models with regular and twisted boundary conditions. We will discuss the $\nu^{(L)}=1$ limit in the subsections below.


\paragraph{Hamiltonian} 

We are interested in the nearest neighbour Hamiltonian $\mc{H}= \sum\mc{H}_{ii+1}$, which is obtained by taking a logarithmic derivative of the transfer matrix. This follows directly from the expansion of the transfer matrix and definition of the Lax operator. By choosing a convenient normalization, the Hamiltonian density is
\eq{
\mc{H}_{ij} := \hbar \frac{(x^+ - \frac{1}{x^+})(x^- - \frac{1}{x^-})}{(x^+ +  x^-)(1  - \frac{1}{x^+ x^-} )} \, \mc{R}_{ji} \, \pa_i \mc{R}_{ij} \Big|_{u_i=u_j=u} , \label{eq:HamDef}
}
where $\pa_i = \frac{\pa}{\pa u_i}$. The parameters $x^\pm$ parametrize the Hamiltonian of the general model for a homogeneous lattice. 

A lengthy but direct computation reveals that the Hamiltonian density can be elegantly presented if we decomposed it into four terms. For this purpose we introduce functions
\spl{
h_1 &= \frac{ x^+ - x^-}{x^+ + x^-} , \qq
h_2 = \frac{1 - x^-  x^+}{x^+ + x^-} , \\
h_\pm &= 1 + \frac{(x^\pm)^{-1}-x^\pm}{\nu^{\pm1} - \nu^{\mp1}} . \label{eq:HamParams}
}
The Hamiltonian density then has the form
\eq{\label{eq:HamGen} 
\mc{H}_{ij} = h_1 \mc{K}_{ij} - 2 h_2 \mc{U}_{ij} - \big (h_2 + \frac{\hbar}{2h_1}\big)\, \mc{V}_{ij} + \frac{\hbar}{4h_1} .
}
The kinetic term $\mc{K}_{ij}$ is 
\eqn{
\mc{K}_{ij} &=\!\! \sum_{\al\ne\bet=\ua\da} \! \big( \ee^{\ii \phi_\al (i-j)}{\rm c}^\dag_{i\al} \ms{c}_{j\al} \mc{N}^{+}_{ij\bet} + \ee^{\ii \phi_\al (j-i)}\ms{c}_{i\al} \ms{c}^\dag_{j\al} \mc{N}^{-}_{ij\bet} \big) , \\[.25em]
\mc{N}^{\pm}_{ij\bet} &= 1 \!- h_\pm (\ms{n}_{i\bet} + \ms{n}_{j\bet} - 1)\! -\! (h_+ + h_-)(\ms{n}_{i\bet}\!-\!1)(\ms{n}_{j\bet} \!-\!1) .
}
The term $\mc{V}_{ij}$ is the usual Hubbard model potential
\eq{ 
\mc{V}_{ij} = (\nmui{i}\!-\!\half) (\nmdi{i}\!-\!\half) +(\nmui{j}\!-\!\half) (\nmdi{j}\!-\!\half)  .
}
The term $\mc{U}_{ij}$ has a simple interpretation in terms of the spin and charge operators. Set
\begin{align}
& \mc{S}_{i,1}^{ch} = \ihalf(\mc{F}_{i,1} \!-\! \mc{E}_{i,1}), \;\; \mc{S}_{i,2}^{ch} = \half(\mc{F}_{i,1} \!+\! \mc{E}_{i,1}), \;\; \mc{S}_{i,3}^{ch} = \minus\half \mc{H}_{i,1} ,\el
& \mc{S}_{i,1}^{sp} = \ihalf(\mc{F}_{i,3} \!-\! \mc{E}_{i,3}), \;\; \mc{S}_{i,2}^{sp} = \half(\mc{F}_{i,3} \!+\! \mc{E}_{i,3}), \;\; \mc{S}_{i,3}^{sp} = \half \mc{H}_{i,3}.\nn
\end{align}
These operators form two copies of the $\alg{su}(2)$ algebra and, for $1\le a,b \le 3$, satisfy $[\mc{S}_{i,a},\mc{S}_{i,b}] = \ii \sum_{1\le c\le 3}\eps_{abc} \mc{S}_{i,c}$, where $\eps_{abc}$ is the Levi-Civita symbol. Recall that the Hamiltonian density of the Heisenberg model, that describes a one--dimensional lattice of spin-$\half$ particles with spin--spin or charge--charge interactions, is
\begin{align}
\mc{H}_{ij}^{sp/ch} := \sum_{1\le a\le 3} \mc{S}_{i,a}^{sp/ch} \mc{S}_{j,a}^{sp/ch}.
\end{align}
This allows us to write the term $\mc{U}_{ij}$ as a difference of the charge and spin Hamiltonian densities
\eq{
\mc{U}_{ij} =  \mathcal{H}^{ch}_{ij} - \mathcal{H}^{sp}_{ij} .
}
All the terms in the Hamiltonian density are individually invariant under the partial particle--hole transformation, $[ P_\da , \mc{K}_{ij} ] = [ P_\da , \mc{U}_{ij} ] = [ P_\da , \mc{V}_{ij} ] = 0$, except the constant term $\frac{1}{4}\hbar/h_1$, which is anti-invariant. 

Set $\phi_\al=\pi/2$ and $\ii\hbar = 4 U$. Use the parametrization \eqref{eq:hub-par} together with the identification $\rm h(\theta) = -\ell$ and $\theta=\mu$. Then the Hamiltonian density \eqref{eq:HamGen} coincides with the one given by equation (12.229) in \cite{book}, up to the sign of the constant term, or in other words, up to the $P_\da$--transformation.

The symmetries of the Hamiltonian density follow directly from the symmetry properties of the oscillator R-matrix. For instance, both the charge and spin $\alg{su}(2)$ symmetries are manifest. This is however not true for the fermionic symmetries.  Let us consider the fermionic intertwining symmetry of $\mc{R}_{ij}$. We have (for clarity, we write the $u_i$ and $u_j$ dependence explicitly)
\spl{ \label{eq:RQ}
&(\mc{Q}_{i}(u_i) \,\nu(u_j) + \mc{Q}_{j}(u_j))\, \mc{R}_{ij}(u_i,u_j) \\
&\qq = \mc{R}_{ij}(u_i,u_j)\, (\mc{Q}_{i} (u_i) + \nu(u_i)\,\mc{Q}_{j}(u_j)) .
}
Expanding this identity to first order in the neighbourhood of $u\equiv u_i=u_j$ and multiplying both sides of it by $\mc{P}_{ij}$, we find 
\spl{
&[ \mc{H}_{ij}(u) , \mc{Q}_{i}(u) + \nu(u)\, \mc{Q}_{j}(u)] \\
& \qq = (\nu(u) - 1) (\pa \mc{Q}_{i})(u) - (\pa\nu)(u)\, \mc{Q}_{j}(u) ,
}
where $\pa= \frac{\pa}{\pa u}$. In other words, we find that even though the oscillator R-matrix has fermionic symmetries, they are not manifest at the level of the Hamiltonian density.


\paragraph{Jordan-Wigner transformation}

To make the double fer\-mionic nature of the model manifest, it is useful to introduce the so-called Jordan--Wigner transformation.
Consider two families of mutually commuting Pauli matrices, $\sigma^{a}_i$ and $\tau^{a}_i$ with $a=\pm,z$ and $1\le i \le L$. The Jordan--Wigner transformation is the map $\ms{JW} : \Osc_{1...L} \to \End V_{4L}$ \mbox{defined by}
\spl{
& \ms{c}_{i,\ua} \mapsto (\sigma^z_1\cdots \sigma^z_{i-1}) \sigma^-_i, \\
& \ms{c}_{i,\da} \mapsto (\sigma^z_1\cdots \sigma^z_L)  (\tau^z_1\cdots \tau^z_{i-1}) \tau^-_i,  \label{eq:JW}
}
 where $V_{4L}=(V^\si_2 \op V^\tau_2)^{\ot L}$ is a $4L$--dimensional vector space equipped with a natural left action of the Pauli matrices $\si_i$ and $\tau_i$ (on the $i^{th}$ copy of two--dimensional spaces $V^\si_2$ and $V^\tau_2$ in $V_{4L}$, respectively). The map $\ms{JW}$ for $\ms{c}^\dag_{i\al}$ is obtained by hermitian conjugating \eqref{eq:JW}.

Applying the Jordan--Wigner transformation to the Hamiltonian \eqref{eq:HamGen}, we find the Hamiltonian for a coupled spin model with kinetic and potential terms given by
\eqn{
\ms{K}_{ii+1} &= (\sigma^+_i\sigma^-_{i+1} \ms{N}^+_{ii+1\tau} + \sigma^-_i\sigma^+_{i+1} \ms{N}^-_{ii+1\tau})  + (\sigma\leftrightarrow\tau) ,\\
\ms{N}^\pm_{ii+1\tau} &=  1 \!-\! \half h_\pm (\tau^z_i \!+\! \tau^z_{i+1}) \!-\! \quarter (h_+ \!+\! h_-)(\tau^z_i \!-\!1)(\tau^z_{i+1} \!-\!1) , \\
\ms{U}_{ii+1} &= \sfrac{1}{8}(\sigma^z_i\tau^z_{i+1}+ \tau^z_i\sigma^z_{i+1}) \\
&\quad - \sfrac{1}{2}(\sigma^+_i\sigma^-_{i+1}-\sigma^-_i\sigma^+_{i+1})
 (\tau^+_i\tau^-_{i+1}-\tau^-_i\tau^+_{i+1})  , \\
\ms{V}_{ii+1} &= \half( \sigma^z_i\tau^z_i +  \sigma^z_{i+1}\tau^z_{i+1} ) ,
}
where $\ms{K}_{ij} = \ms{JW}(\mc{K}_{ij})$ and similarly for the other terms.


\section{A, B and EKS models}


There are four limits of the general model, that correspond to special points on the elliptic torus parametrized by Zhukovsky variables $x^\pm$ (see explanation in Section 2.2 in \cite{FQ}). In this section we discuss three limits that correspond to the A and B models of Frolov--Quinn \cite{FQ}, and the Essler--Korepin--Schoutens (EKS) model \cite{EKS1}.  We also describe the Yangian symmetries of the A and B models. The fourth case, the Hubbard model limit, will be considered separately in Section 6. 

\paragraph{The A model}

The A model is obtained by setting 
\eq{
\phi_\al = 0, \qq x^+ = -\frac{1}{x^-} = \ee^{\ii \theta}.
}
The Hamiltonian parameters \eqref{eq:HamParams} in this limit become
\eq{
h_1 = \ii \cot \theta , \qu
h_2 = \ii \csc \theta , \qu
h_\pm = 1 \pm \tan \theta .
}
The symmetries of the A model have a non-trivial braiding and consequently the fermionic symmetries of the R-matrix do not lead to manifest symmetries of the Hamiltonian. However, both spin and charge operators are symmetries of the Hamiltonian
\eq{
[\mc{H}, \mc{J}_{a}] = 0, \qu \mc{J} \in \{\mc{E} ,\mc{F} ,\mc{H} \}, \qu a=1,3.
}
As in the Hubbard model, these $\alg{su}(2)$ symmetries can be enhanced to Yangian symmetries. Remarkably, the Yangian symmetries of the A model follow from the ones of the general model displayed in \eqref{eq:Yang} by making use of the fermionic operators. One can check that
\eq{
\hat{\mc{J}}^{A}_{a} : =  \hat{\mc{J}}_a +2 \ii \sin \theta\,  \sum_{\alpha,\beta}[\mc{Q}_{\alpha\beta},\mc{J}_a]\mc{Q}^\dag_{\alpha\beta}+
[\mc{Q}^\dag_{\alpha\beta},\mc{J}_a]\mc{Q}_{\alpha\beta}
}
commutes with the Hamiltonian of the A model. Moreover, this model also has a secret symmetry given by
\eq{
\hat{\mc{B}}^{A} : =  \hat{\mc{B}} + 2\ii(1+\cos^2 \theta)\csc \theta\, \sum_{\alpha,\beta=\ua,\da}\mc{Q}_{\alpha\beta}\mc{Q}^\dag_{\alpha\beta}.
}
%


\paragraph{The B model}

The B model corresponds to the case
\eq{\label{eq:paramB}
\phi_\al = -\half \pi, \qq x^+ = -x^- = \ii \cos \theta. 
}
The general Hamiltonian diverges in this limit and has to be normalized by multiplying by $(x^+ + x^-)$ to yield a finite result. By doing so we find
\eq{
h_1 = 2 \ii \cos \theta , \qu
h_2 = \sin^2 \theta , \qu
h_\pm = -2 \sec \theta \sin^4 \frac{\theta}{2} .
}
Note that the term $\dfrac{\hbar}{2h_1} \mc{V}_{ij}$ in \eqref{eq:HamGen} vanishes in this limit.

All the symmetries of the general model are present in the B model as well. However, the parameters in the symmetry generators are different from the ones in \eqref{eq:paramB}. In particular, we have to choose different values of the phase factors and identify $x^+ \sim 1/x^+$, so that
\eq{
\phi_\ua=-\phi_\da = \half \pi, \qq x^+ = -x^- = -\ii \csc \theta.
}
By evaluating representation parameters in this way, we find that Yangian operators in \eqref{eq:Yang} are symmetries of the B model. Furthermore, we reproduce the fermionic operators $\bf{Q}_{\al\si},\bf{Q}^\dag_{\al\si}$ given in Section 2.3 in \cite{FQ} by setting the phases $\phi_\ua=\half\pi$ for $\mc{Q}_{\al\al},\mc{Q}^\dag_{\al\al}$ and $\phi_\ua=-\half\pi$ for $\mc{Q}_{\al\neq\beta},\mc{Q}^\dag_{\al\neq\beta}$, respectively. Finally, the B model exhibits a secret symmetry upon setting $\phi_\al=0$ in \eqref{eq:Secret}.


\paragraph{EKS model}

The Essler--Korepin--Schoutens model is contained in the general model for a specific value of its coupling constants. In particular, this corresponds to the case $U=h=\mu=0$ of equation (6) in \cite{EKS1}. This limit is obtained by choosing a different parametrization of $x^\pm$. Instead of \eqref{x(u)}, we need to set
\eq{
x(u_i) = \frac{1}{2}\left( u_i + \sqrt{u^2_i-4}\right),
\qu x^\pm_i = x(u_i\pm\half \hbar) . \label{eq:EKS-x(u)}
}
Then the above case of the EKS model is obtained by setting $\phi_\al = 0$ and taking the $u \rightarrow\infty$ limit. As in the B model, we need to normalize the general Hamiltonian by an additional power of $u^{-1}$ to obtain a finite result. This gives
\eq{
h_1 = 0, \qu h_2 = -\half ,\qu  h_1 h_\pm = \mp \half , \qu \frac{\hbar}{2h_1}=1 ,
}
which means that the kinetic term $\mc{K}_{ij}$ does not vanish, rather its $\mc{N}^{\pm}_{ij\bet}$ terms are now defined by  
\eq{
\mc{N}^{\pm}_{ij\bet} = \pm\half(\ms{n}_{i\bet} + \ms{n}_{j\bet} - 1) .
}
For the EKS model, all the fermionic symmetries are manifest since $\nu_i=1$ in this limit (i.e.\@ the last term in \eqref{eq:[t,Q]} vanishes). This model actually has a manifest $\mf{u}(2|2)$ symmetry. We note that the auxiliary space of the EKS model is different from ours (and that of the A and B models). One also needs to take a similar limit of the auxiliary parameters $x^\pm_\mu$.


\section{The Hubbard model}

In the remaining two sections we consider the Hubbard model limit of the general model. For ease of notation, we index with $\infty$ all the relevant operators in this limit. 


\paragraph{Lax operator}

The Hubbard Lax operator is obtained by setting $\phi_\al=\pi/2$ and taking the limit
\eq{
\mc{L}_{\mu i}^{\infty}:=\lim_{u_i\to\infty}\mc{L}_{\mu i} \,\big|_{\phi_\al=\pi/2} \label{eq:Lax-Hub}
}
of the quantum spectral parameter $u_i$. In this limit the spin chain is homogeneous. We find that $\lim_{u_i\to\infty} p^\pm_{\mu i}=\lim_{u_i\to\infty} q^\pm_{\mu i}=0$, thus $P^\infty_{\mu i}=Q^\infty_{\mu i}=0$ and $\mc{L}_{\mu i}^{\infty}$ becomes of the standard form. The coefficients of the matrix elements are
$$
(f_\mu)^2 = \frac{\nu_\mu - 1}{\nu_\mu + 1}, \;\; (k^+_\mu)^2 = 1 - \nu_\mu^{-2}, \;\; (k^-_\mu)^2 = \frac{x^+_\mu - x^-_\mu}{\nu_{\mu}},
$$
where we have suppressed the quantum space index $i$. In the standard parametrization \eqref{eq:hub-par}, $f_\mu = \ee^{i \theta}$, $k^+_\mu = \sec \theta$ and $k^-_\mu = \ee^{{\rm h}(\theta)} \sec\theta$.  The inverse of the Hubbard Lax operator is obtained by substituting $x^\pm_\mu \to 1/x^\mp_\mu$, or, equivalently, by $\rm h(\theta) \to - \rm h(\theta)$.

The Hubbard Lax operator satisfies the spin and charge symmetries \eqref{eq:[L,J]}, However, the fermionic symmetry \eqref{eq:[L,Q]} does not have its counterpart in the Hubbard limit, since $\nu_i \to \infty$ in this limit.


\paragraph{Transfer matrix} 

The Hubbard monodromy matrix $\mc{T}_\mu^\infty$ and transfer matrix $\tau_\mu^\infty$ are obtained by taking the limit \eqref{eq:Lax-Hub} of all quantum spectral parameters $u_i$ and setting $\phi_\al=\pi/2$,
\eq{
\mc{T}_\mu^\infty := \lim_{u_i \to \infty} \mc{T}_\mu \,\big|_{\phi_\al=\pi/2}, \qu
\tau_\mu^\infty := \Str_\mu \mc{T}_\mu^\infty .
}
They satisfy the spin and charge symmetry relations \eqref{eq:[T,J]}, althought the symmetries \eqref{eq:[T,Q]} and \eqref{eq:[t,Q]} do not survive.


\paragraph{Hamiltonian}

The Hubbard Hamiltonian is obtained by taking the $u\to\infty$ limit of the Hamiltonian density \eqref{eq:HamGen} and setting $\phi_\al=\pi/2$, 
\eq{
\mc{H}_{ij}^\infty := \lim_{u\to \infty} \mc{H}_{ij} \big|_{\phi_\al=\pi/2} . 
}
In this limit the constants $h_\pm,h_2$ tend to zero, while $h_1$ tends to unity. Therefore $\mc{N}_{ij\bet}^{\pm\infty} = 1$ and we obtain the usual Hamiltonian density of the Hubbard model, $\mc{H}^\infty_{ij} = \mc{K}_{ij}^\infty - \half\hbar \mc{V}_{ij}^\infty$, given by
\eq{
\mc{K}_{ij}^\infty = \sum_{\al=\ua\da} \! \big( \ii^{i-j}{\rm c}^\dag_{i\al} \ms{c}_{j\al} + \ii^{j-i}\ms{c}_{i\al} \ms{c}^\dag_{j\al} \big) , \qu
\mc{V}_{ij}^\infty = \mc{V}_{ij}  . \label{eq:KV-hub} 
}
Another possibility is to set $\phi_\al=-\pi/2$, which is equivalent to replacing $\ii$ with $-\ii$ in \eqref{eq:KV-hub}. This effectively gives a Hubbard model with opposite coupling constant, $-\hbar$.


\section{A shortcut to the Hubbard Hamiltonian and the Uglov--Korepin Yangian}

Yangian operators \eqref{eq:Yang} and \eqref{eq:Secret} are symmetries of the R-matrix, but not of the Hamiltonian \eqref{eq:HamGen}. However they become symmetries of the Hubbard Hamiltonian in the $u\to\infty$ limit. In particular, spin and charge Yangian symmetries specialize to the Uglov--Korepin Yangian obtained in \cite{UK}, and the secret symmetry specializes to the Hubbard Hamiltonian. We will demonstrate this specialization for the spin Yangian operators. The specialization of the charge Yangian operators is similar. 

Let $\mc{J}\in\{\mc{E},\mc{F},\mc{H}\}$. The Uglov--Korepin Yangian operators in (\cite{UK}, Statement 1) in our notation can be written~as
\eq{
\wh{\mc{J}}^\infty_3 = \sum_{1\le i< L} (\mc{J}^{(+1)}_{i,3} + \mc{J}^{(-1)}_{i+1,3}) + \frac{\hbar}{2} \sum_{1\le i<j \le L} \big [ \mc{J}_{i,3} , \mc{Z}_{ij} \big] , 
}
where the non-local spin symmetry operators are defined by
\spl{\label{eq:KUgens}
\mc{E}^{(j)}_{i,3} &:= \ee^{\ii(\phi_\ua i - \phi_\da (i+j))} \ms{c}^\dag_{i\ua} \ms{c}_{i+j,\da} \\
\mc{F}^{(j)}_{i,3} &:= \ee^{\ii(\phi_\da i - \phi_\ua (i+j))} \ms{c}^\dag_{i\da} \ms{c}_{i+j,\ua}, \\
\mc{H}^{(j)}_{i,3} &:= \ee^{\ii \phi_\da j} \ms{c}^\dag_{i\da} \ms{c}_{i+j,\da} - \ee^{\ii \phi_\ua j} \ms{c}^\dag_{i\ua} \ms{c}_{i+j,\ua} ,
}
and we have assumed that $\phi_\al=\pi/2$. A direct computation then shows that $\wh{\mc{J}}^\infty_3$ is a symmetry of the Hubbard Hamiltonian on an infinite interval.

The following linear combination of the spin Yangian generators and the quadratic combination of fermionic operators in $\rho(\alg{g})$ specializes to the Uglov--Korepin Yangian operators
\eq{ \label{LimUK}
\lim_{u\to\infty} \bigg( \wh{\mc{J}}_3 - \frac{\hbar}{2} \sum_{\al,\bet=\ua,\da} \big[\mc{J}_{3} , \mc{Q}_{\al\bet}\big]\mc{Q}^\dag_{\al\bet}  \bigg) = - \wh{\mc{J}}^\infty_{3} . 
}
Moreover, the specialization
\eq{ \label{LimH}
\!\!\lim_{u\to\infty} \!\bigg( \si(\wh{\mc{B}}) - \frac{\hbar}{2} \sum_{\al,\bet=\ua,\da} \si(\mc{Q}_{\al\bet}\mc{Q}^\dag_{\al\bet}) + L(u-\sfrac{\hbar}{4})\bigg) =\minus \mc{H} \!
}
relates the secret symmetry to the Hubbard Hamiltonian. 

Let us now prove these specializations. First, note that taking the $u\to\infty$ limit of \eqref{eq:QQd} and keeping the leading terms only, we have
\spl{
\mc{Q}_{i\ua\ua}^\infty &=  \tfrac{1}{\sqrt{\hbar}} \ee^{-\ii\phi_\ua i} u\, \cui{i}, && \mc{Q}^{\dag\infty}_{i\ua\ua} = \tfrac{1}{\sqrt{\hbar}} \ee^{\ii\phi_\ua i} \cdui{i}, \\
\mc{Q}_{i\ua\da}^\infty &= \tfrac{1}{\sqrt{\hbar}}\ee^{\ii\phi_\da i} u\, \cddi{i}, \qu && \mc{Q}^{\dag\infty}_{i\ua\da} = \tfrac{1}{\sqrt{\hbar}} \ee^{-\ii\phi_\da i} \cdi{i}, \\
\mc{Q}_{i\da\ua}^\infty &= -\tfrac{1}{\sqrt{\hbar}} \ee^{-\ii\phi_\da i} u\, \cdi{i}, && \mc{Q}^{\dag\infty}_{i\da\ua} = -\tfrac{1}{\sqrt{\hbar}} \ee^{\ii\phi_\da i} \cddi{i}, \\
\mc{Q}_{i\da\da}^\infty &= \tfrac{1}{\sqrt{\hbar}} \ee^{\ii\phi_\ua i} u\, \cdui{i}, && \mc{Q}^{\dag\infty}_{i\da\da} = \tfrac{1}{\sqrt{\hbar}} \ee^{-\ii\phi_\ua i} \cui{i} .
}
Upon setting
\eq{
C_{ij} = \sum_{\al=\ua,\da}(\ee^{\ii \phi_\al(i-j)}\ms{c}^\dag_{i\al}\ms{c}_{j\al} + \ee^{\ii \phi_\al(j-i)} \ms{c}_{i\al}\ms{c}^\dag_{j\al})
}
and keeping the leading terms only, we have
\eqa{
\mc{W}^{\pm\infty}_{ij} = \frac{1}{\hbar} ( u^{j-i+1} \pm u^{i-j+1} ) C_{ij}, \label{eq:limW}\\
\sum_{\al,\bet=\ua,\da} \mc{Q}^\infty_{\al\bet} \mc{Q}^{\dag\infty}_{\al\bet} = \frac{1}{\hbar}\sum_{1\le i,j\le L} u^{i-j+1}C_{ij} . \label{eq:limQQd}
}
Notice that $C_{ii}=2$, $C_{ij}=-C_{ji}$ for $i\ne j$ and 
$
[\mc{J}_{i,3},C_{ij}]= \mc{J}^{(j-i)}_{i,3} + \mc{J}^{(i-j)}_{j,3}.
$
Hence
\eqn{
A & := \sum_{1\le i<j\le L} \big[ \mc{J}_{i,3} , \mc{W}^{-\infty}_{ij} ] - \sum_{\al,\bet=\ua,\da} \big[\mc{J}_3 , \mc{Q}^\infty_{\al\bet} \big] \mc{Q}^{\dag\infty}_{\al\bet} \el
& \, = -\frac{2}{\hbar}\sum_{1\le i<j\le L} u^{i-j+1} [\mc{J}_{i,3},C_{ij}] - \frac{2u}{\hbar}\sum_{1\le i \le L} \mc{J}_{i,3} .
}
In the $u\to\infty$ limit only the $j=i+1$ terms contribute in the sum above. Thus
$$
\lim_{u\to \infty} \Bigg( \sum_{1\le i\le L} \!u\, \mc{J}_{i,3} + \frac{\hbar}{2} A \Bigg) = - \sum_{1\le i < L}  (\mc{J}^{(+1)}_{i,3} + \mc{J}^{(-1)}_{i+1,3}) ,
$$
which, combined with \eqref{eq:Yang}, gives \eqref{LimUK}. 

Next, we demonstrate the specialization \eqref{LimH}. Observe that
\eqa{
\sum_{1\le i<j\le L} \si(\mc{W}^{+\infty}_{ij}) &= - \sum_{1\le i<j\le L} \mc{W}^{+\infty}_{ij}, \\
\sum_{\al,\bet=\ua,\da} \si(\mc{Q}^\infty_{\al\bet} \mc{Q}^{\dag\infty}_{\al\bet}) &= \frac{1}{\hbar}\sum_{1\le i,j\le L} u^{j-i+1}C_{ij} .
}
Therefore
\eqn{
B & := \sum_{1\le i<j\le L} \si(\mc{W}^{+\infty}_{ij}) + \sum_{\al,\bet=\ua,\da} \si(\mc{Q}^\infty_{\al\bet} \mc{Q}^{\dag\infty}_{\al\bet}) \el
& \, = -\frac{1}{\hbar}\sum_{1\le i\le j\le L} (2-3\del_{ij}) u^{i-j+1} C_{ij}.
}
and
$$
\lim_{u\to \infty} \frac{\hbar}{2} B  = \sum_{1\le i < L}\sum_{\al=\ua,\da} \ii(\ms{c}^\dag_{i\al}\ms{c}_{i+1,\al}+\ms{c}^\dag_{i+1,\al}\ms{c}_{i\al}) + L u ,
$$
which, combined with the identity
$$
\lim_{u\to\infty} \frac{u}{ad+bc}\,(\mc{H}_{i,3})^2 = -\hbar (\nmui{i}-\half)(\nmdi{i}-\half) + \frac{\hbar}{4}
$$
and \eqref{eq:Secret} gives \eqref{LimH}, as required. The key feature of this specialization is the braided non-local terms $\mc{W}^\pm_{ij}$ of the Yangian and secret symmetry operators. In the $u\to\infty$ limit these non-local terms specialize to the nearest-neighbor terms that describe the hopping terms in the Hamiltonian or the hopping-like terms in \eqref{eq:KUgens}. The Hamiltonian is related to the `opposite' secret symmetry $\si(\hat{\mc{B}})$ because of the definition \eqref{eq:HamDef} and the fact that $\mc{R}_{ji}(u,u) = \mc{P}_{ji}$.


\section{Conclusions}

In this letter we have discussed the algebraic structure of a general integrable lattice model described by the centrally extended $\mf{su}(2|2)$ superalgebra. This superalgebra is a symmetry of Beisert's AdS/CFT worldsheet R-matrix and Shastry's R-matrix, and gives rise to a one--dimensional Hubbard--Shastry type model. Quantum spaces of this model are parametrized by quantum spectral parameters and are identified with fundamental modules of this model's superalgebra. Despite the fact that the lattice and the R-matrix exhibit both bosonic and fermionic symmetries, only bosonic symmetries are manifest in the Hamiltonian. The fermionic symmetries enter only via certain non-abelian identities.

The general model contains the A and B models introduced in \cite{FQ}, and a special case of the EKS model \cite{EKS1} as certain specializations. Starting with the Yangian symmetries of the general model, we found Yangian and secret symmetries of the A and B models. 

Finally, we identified the limit in which the general model specializes to the one--dimensional Hubbard model. We have shown that spin and charge Yangian symmetries of the general model in this limit specialize to the Uglov--Korepin Yangian operators. Remarkably, we find a purely algebraic interpretation of the Hubbard Hamiltonian as a specialization of the secret symmetry.

\smallskip 

\noindent {\it Acknowledgements.} The authors would like to thank A. Prinsloo and A. Torrielli for useful discussions and comments. M.d.L. was supported by FNU through grant number DFF-1323-00082. V.R. thanks the Engineering and Physical Sciences Research Council (EPSRC) of the United Kingdom for the Postdoctoral Fellowship under the grant EP/K031805/1 `New algebraic structures inspired by gauge/gravity dualities'.


\appendix

\section{}  \label{App:A}

\noindent The Cartan matrix and the distinguished Dynkin diagram of the centrally extended $\mf{su}(2|2)$ superalgebra are
\eq{
A = \left(\begin{array}{ccc} 2 & -1 & 0 \\ -1 & 0 & 1 \\ 0 & 1 & -2 \end{array}\right)  \qq 
\begin{tikzpicture}[baseline=-7pt]  
\draw [thick] (0.15,0) -- (.45,0);
\draw [thick] (0.75,0) -- (1.05,0);
\draw [thick] (0.5,0.1) -- (.7,-0.1);
\draw [thick] (0.5,-0.1) -- (.7,0.1);
\draw [thick] (0,0) circle [radius=0.15];
\draw [thick] (.6,0) circle [radius=0.15];
\draw [thick] (1.2,0) circle [radius=0.15];
\node [below] at (0,-0.2) {1};
\node [below] at (0.6,-0.2) {2};
\node [below] at (1.2,-0.2) {3};
\end{tikzpicture} 
}

\noindent The matrix $R(u_i,u_j)$ has the form
\eqa{
R&(u_i,u_j) = \el
&= ((1 -  s^{(4)}_{ij}) \textstyle\sum\limits_{\bar a=1,\, \bar b=0} 
\!+\, (1 -  s^{(5)}_{ij}) \textstyle\sum\limits_{\bar a=0,\, \bar b=1} )\, E_{aa}\ot E_{bb}  \el
& + (s^{+}_{ij} -  s^{-}_{ij})  \textstyle\sum\limits_{\bar{a}+\bar{b}=1} \!(\minus)^{\bar{a}} E_{ab}\ot E_{ba} + s^{(3)}_{ij} \!\! \textstyle\sum\limits_{\bar a = \bar b = 1} \! E_{aa} \ot E_{bb} \el
& + ( s^{(2)}_{ij} \!\textstyle\sum\limits_{\substack{a\neq b\\\bar{a}=\bar{b}=0}} \!+\, s^{(3)}_{ij} s^{(2)}_{ji} \!\textstyle\sum\limits_{\substack{a\neq b\\\bar{a}=\bar{b}=1}} ) (E_{aa}\ot E_{bb} - E_{ab}\ot E_{ba}) \el
& + (s^{+}_{ij} + s^{-}_{ij}) \textstyle\sum\limits_{\substack{\bar a = \bar c,\, \bar b = \bar d\\ a\ne b \ne c \ne d}}(\minus)^{\overline{a \scalebox{0.6}[.8]{$-$} b} \scalebox{0.6}[.8]{$-$} \bar a} E_{ab}\ot E_{cd} ,
}
with the Boltzmann weights given in \eqref{eq:BolzRmat}.

\medskip


\end{document}